\documentclass[aps,prb,reprint]{revtex4-1}
\usepackage{blindtext}
\usepackage{appendix}
\usepackage{graphicx}
\usepackage{epsfig} 
\usepackage{caption}
\usepackage{subfloat}
\usepackage[justification=raggedright, font=small, labelfont=bf]{caption}
\usepackage[dvipsnames]{xcolor}
\usepackage[T1]{fontenc}
\usepackage{wrapfig}
\usepackage[english]{babel}
\usepackage{mathrsfs} 
\usepackage{amsmath}
\usepackage{amsfonts}
\usepackage{sidecap}
\usepackage{bbm}
\usepackage[multidot]{grffile}
\usepackage{subfig}
\usepackage{multirow}
\usepackage{url}
\usepackage[english]{babel}
\usepackage{soul}

\newcommand{\be}{\begin{equation}}
\newcommand{\ee}{\end{equation}}

\begin{document}
\title{Nanoscopic time crystal obtained by nonergodic spin dynamics}
\author{Carla Lupo}
\affiliation{King's College London, Theory and Simulation of Condensed Matter, The Strand, WC2R 2LS London, UK}
\author{Cedric Weber}
\affiliation{King's College London, Theory and Simulation of Condensed Matter, The Strand, WC2R 2LS London, UK}

\begin{abstract} 
We study the far-from-equilibrium properties of nanoscopic classical spin systems. In particular, we focus on the interplay between lattice vibrations and magnetic frustrations induced by surface effects in antiferromagnets. We use an extended Monte Carlo simulations which treats both the ionic degrees of freedom and spin variables on the same footing, via a Heisenberg-Lennard-Jones Hamiltonian with a spin-lattice coupling. The interplay of the local ordered magnetic moments and the lattice dynamics provides, at zero temperature, a structural phase diagram characterizing the magnetic order in two different antiferromagnetic nanoclusters. At non zero temperature, the competition between spins and the ionic vibrations considerably affects the magnetization of these systems. Next, we explore the dynamical response of the antiferromagnetic structures subjected to an initial ferromagnetic quench by solving the stochastic Landau-Lifshitz-Gilbert equation at finite temperature. The dynamics reveals a non trivial structural induced behavior in the spin relaxation with a concomitant memory of the initially applied ferromagnetic quench.  These observations of long-lived non-thermal states could open new avenues in nano technology.
\end{abstract}

\maketitle
\section{Introduction}
Many-body systems comprise a wide range of systems, from simple metals, organic molecules, all the way up to cells. While their physics can be extremely rich, this complexity is however often  irrelevant, as such systems will typically - at equilibrium - thermalise, a process through which most information about their preparation history and their initial state is lost \cite{Srednicki_1994, Rigol_2008}. This behavior is typical for ergodic systems in the thermodynamic limit and allows us to calculate physical observables and make predictions that can be measured and tested.
However, ergodicity can be broken out-of-equilibrium  \cite{Garrahan_2018}, in particular by inducing non-thermal states that keep a memory of their initial condition for long times.
Those peculiar behaviors have been explored in novel non-equilibrium phases of matter, which includes Floquet symmetry protected topological phases  \cite{Potter_2016,Sentef_2015}
and time crystals \textcolor{black}{\cite{Wilczek_2012,Shapere_2012, Watanabe_2015}},\cite{Patrick_2013, Zhang_2017,Sacha_2018}.
In particular, out-of-equilibrium quantum materials  have been extensively studied theoretically, with a range of quantum approaches, such as the Floquet dynamical mean-field theory \cite{Aoki_2014}  used for electron-spin systems \cite{Mentink_2015, Mikhaylovskiy_2015} and  non-equilibrium Green's function techniques for electron-phonon dynamics \cite{Schuler_2016, Sakkinen_2015}.
While bulk systems have been extensively investigated, nanoscopic systems remain open to questions. 
Recent progress in nano-engineering, such as the design of single atom arrays as memory devices \cite{Yane_2017}, has opened new possibilities to explore quantum states in systems where thermalisation is not obtained, in particular for quenched and periodically-driven systems.
Indeed, it has been observed that finite size effects allow these systems to keep a much better local memory of their initial conditions \cite{Loth_2012}. 
Experimental advances in manipulation and switching of the magnetization - possible even at the femtosecond level \cite{Bigot_2009} - have triggered new studies in spin dynamics \cite{Dutta_2018} towards the microscopic understanding of emerging relaxation time scales and the discovery of new suitable candidates for magnetic-logic building block \cite{Dutta_2018}.
Particular attention has been paid when antiferromagnetic order is involved \cite{Baltz_2018,Vasilakaki_2015} : in these systems the contribution of the uncompensated spins at the surface triggers exotic phenomena which are not analitycally tractable and whose precise understanding is still unknown. To that end, new mechanisms for slow relaxation and non-ergodicity  have potential implications for the design and control of novel quantum non-equilibrium materials and devices. Atomistic electron-spin models have proved to be a powerful approach to model ultrafast magnetization dynamics.\\ 
In this work, we propose a study of the dynamics of a typical many-body nanoscopic system in which the relaxation processes involve both the structure and magnetic moments.
We consider the quenched and driven classical spins, in the presence of  long-range magnetic interactions coupled to the lattice dynamics.  We first focus on the interplay of the local ordered magnetic moments and the lattice dynamics at finite temperature. 
Next, we extend the equilibrium calculations to a ferromagnetic quench. We go on to study the equilibrium and out-of-equilibrium properties using a combination of Monte Carlo simulations  \cite{Chen_1993, Holm_1993} and atomistic spin dynamics \cite{Ma_Dudarev_2017, Tranchida_2018} through the numerical implementation of the Landau-Lifshitz-Gilbert \cite{Brown_1963} equation extended to deal with finite temperatures within a Langevin dynamics framework. 
\\
\section{Equilibrium Properties of Magnetic Nanoclusters}
We consider a cluster in a spherical shape and a simple-cubic structure with unitary lattice constant. 
The total lattice-spin Hamiltonian of the cluster reads as follows:
\be\label{eq:hamiltonian_H_LJ}
H=-\sum_{i,j} J(d_{ij}) \mathbf S_i\cdot \mathbf S_j+K \sum_{i,j} \left[\left(\frac{d^0}{ d_{ij}}\right)^{12}-2\left(\frac{d^0}{d_{ij}}\right)^{6}\right]
\ee
where $\mathbf S_i$ are $O(3)$ unitary $\vert\mathbf S\vert=1$ spins ruled by a Heisenberg hamiltonian with an inhomogeneous super exchange coupling $J(d_{ij})=J_0\,e^{-\alpha d_{ij}}$ where $J_0$ is the antiferromagnetic nearest neighbor interaction. The exchange interaction function $J(d_{ij})$ can be tuned from \textit{ab initio} calculation or experimental results \cite{Ma_Dudarev_2017, Weber_Becca_Mila_2005}. 

\begin{figure}[!t]
\begin{center}
\includegraphics[width=0.5\textwidth,height=0.35\textwidth]{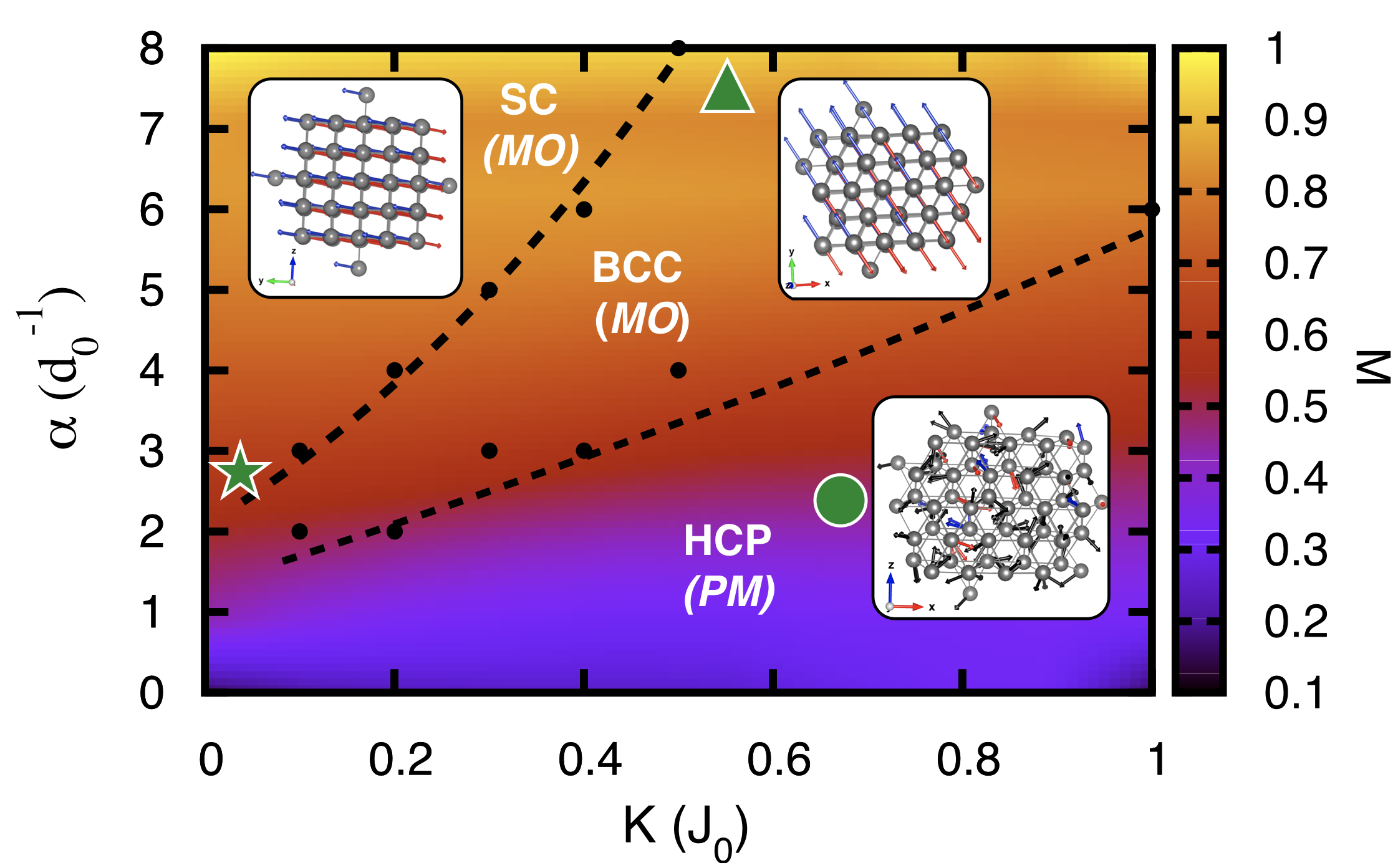}
\caption{(colors online) Zero temperature structural phase diagrams with respect to the interaction range $\alpha$ and the Lennard-Jones potential $K$. Dashed lines are guides to the eyes and distinguish three different structural regions:
simple-cubic (SC, e.g. $\star$ LaTiO$_3$ ), body-centered-cubic (BCC, e.g. $\bigtriangleup$ MnO ) and hexagonal-closed-packed  (HCP, e.g. $\bigcirc$ Ni ).   Shaded colours indicate the presence of antiferromagnetic order (MO). The arrows represent one possible spin configuration for the collinear magnetic order (at $\mathbf Q=(\pi,\pi,\pi)/a$ for SC and $\mathbf Q=(0,0,2\pi)/a$) for BCC and magnetic disorder for the HCP. The system in analysis has 123 sites. }
\label{fig:GS_phase_diagram}
\end{center}
\end{figure}

The $\alpha$ parameter scales the range of the interaction: as $\alpha$ decreases the interaction becomes long range. 
The lattice deformation is constrained by a Lennard-Jones (LJ) potential with a normalised unit distance ($d_0=1$). We use extended Monte Carlo Metropolis calculations with both spins and ionic displacements update \cite{Olive_1986,Miyatake_1986,Weber_Becca_Mila_2005, Weber_Capriotti_2003}.

The competition between the Heisenberg term and the Lennard-Jones potential, respectively tuned by the $\alpha$ and $K$ parameters introduced in Eq.(\ref{eq:hamiltonian_H_LJ}), stabilizes three different ground state structures (Fig \ref{fig:GS_phase_diagram}): simple-cubic (SC), body-centered-cubic (BCC) and hexagonal-close-packed (HCP). These structures are obtained  if the system optimizes either the super-exchange potential or the ionic interactions (LJ). When the dominant contribution to the energy is the exchange term, the structure is SC, with a concomitant lattice compression. Indeed, the SC structure accommodates the largest number of non-frustrated antiferromagnetic bonds in three dimensions. For larger $K$, we obtain a competition between the exchange and LJ terms. This drives a structural transition towards a BCC structure, which increases the coordination number, albeit retaining an ordered spin structure with pitch vector ($\mathbf q=(0,0,2\pi/a)$).  These two cases show magnetic order (MO). For $ K>>1$ the leading contribution comes from the LJ potential, and the system increases further its coordination number losing the magnetic order in favour of the more compact HCP structure. Under the assumption of large magnetic moment, paradigmatic systems for the model above are represented by antiferromagnetic SC structures such as perovskites (e.g. LaTiO$_3$ \cite{Mellouhi_2013, Luan_2015,Weng_2015}), BCC nanocluster such as \textcolor{black}{Mn} \textcolor{black}{\cite{Briere_2002}},\cite{ Duarte_2014, Cardias_2017}  and HCP Ni nanoparticle \cite{Garcia_2011, Filippova_2015, Tranchida_2018}. (Symbols in Fig1 identify the different materials with parameters  listed in Table \ref{tab:table_LJ_H}).

\begin{figure}[!b]
\begin{center}
\includegraphics[width=0.5\textwidth,height=0.35\textwidth]{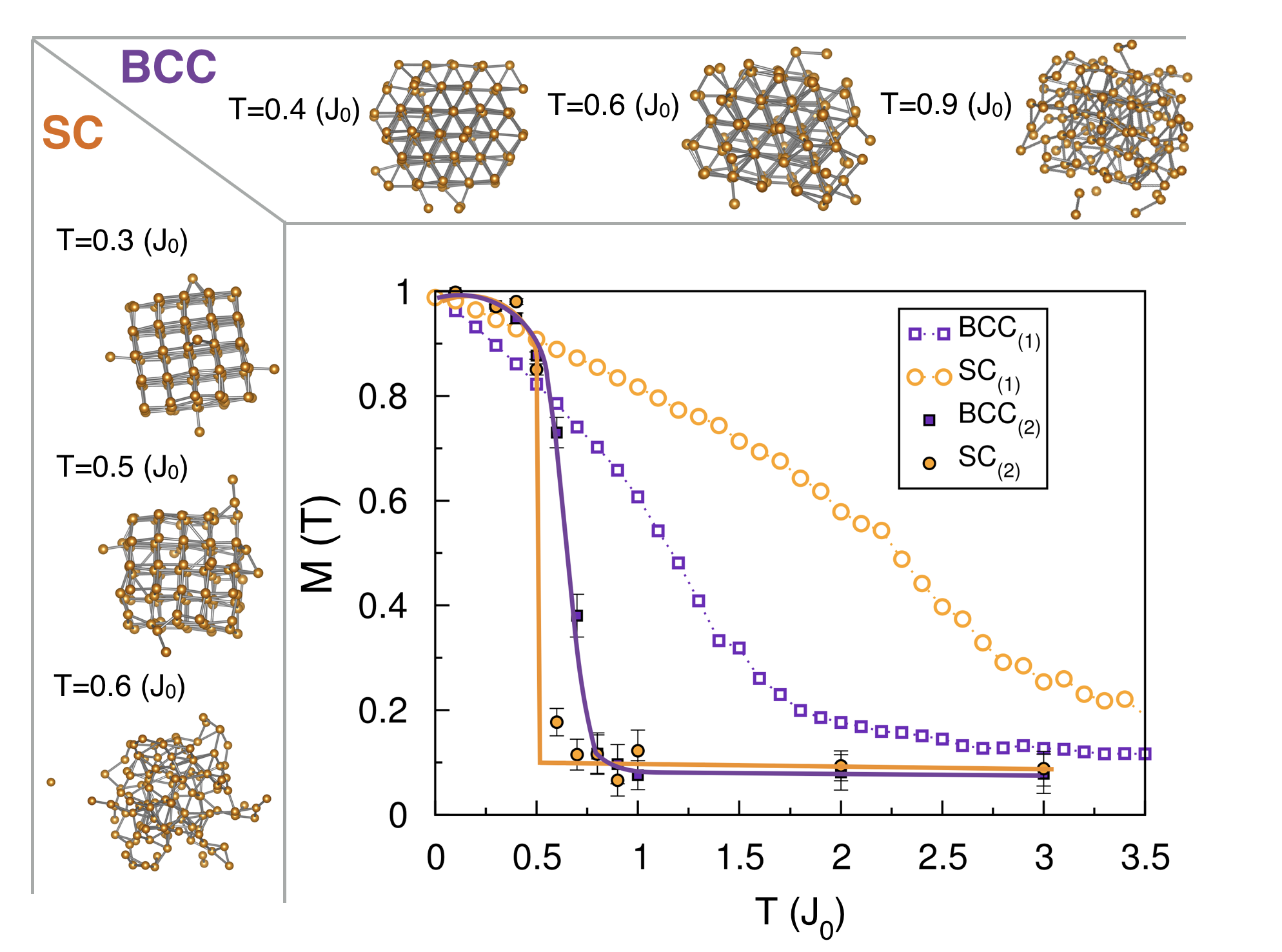}
\caption{(colors online) Temperature dependent behavior of the magnetic order parameter obtained running equilibrium simulations with (sim$_{(2)}$) and without (sim$_{(1)}$) considering the coupling of the spins with the lattice deformation. The two structures considered are SC (orange) and BCC (violet). Top and left panel show the structures obtained at finite temperature.} 
\label{fig:mu_HB_vs_S+L_picture}
\end{center}
\end{figure}

To summarize, we note that the spin-exchange and Lennard-Jones terms compete in a non-trivial fashion in  the presence of antiferromagnetic interaction: on the one hand, the Lennard-Jones term favors the most compact structure, e.g. the HCP, albeit this phase is magnetically frustrated, and on the other hand the optimal phase to accommodate anti-ferromagnetism in three dimensions is the regular cubic lattice. This leads to a very rich phase diagram at zero temperature, as mentioned above, but also induces non-trivial effects at finite temperature (Fig \ref{fig:mu_HB_vs_S+L_picture}). We now turn to finite temperature calculations of the structures with zero temperature magnetic order (SC and BCC in Fig \ref{fig:GS_phase_diagram}). In absence of the ionic motion, the magnetic order parameter of the SC phase (SC$_{(1)}$) is stable until $T \approx 3[J_0]$, whereas for frustrated BCC phase (BCC$_{(1)}$) we find a less stable magnetic phase (magnetic for $T<1.8[J_0]$). However, once both the spin and ionic potential are considered (SC$_{(2)}$, BCC$_{(2)}$), we observed that both structures are magnetic for $T < 0.6[J_0]$, but the processes leading to the paramagnetic phase are very different. For the SC phase, we observe that the drop in magnetization is concomitant with a loss of the structural properties, as the lattice starts melting at $T \approx 0.6[J_0]$ (as shown in the left  panel), whereas the BCC structure survives above this temperature (as shown in the top panel), and undergoes a magnetic transition towards the paramagnetic phase due to magnetic fluctuations. The non zero magnetization value at high temperature is due to the finite size of the system. We observe that the values of M of SC$_{(1)}$, BCC$_{(1)} $at $T=5[J_0]$  are different due to different contribution of the surface spins.  \\

\section{Dynamical response to external fields and memory effects}
We now turn to the discussion of the time evolution after a ferromagnetic quench. We focus here on the time evolution of the system at low enough temperature (far from the melting temperature), where the structure is weakly dependent on temperature and can be considered  fixed. We calculate the dynamical magnetic properties of the nano-structure far from equilibrium, with the Landau-Lifshitz-Gilbert formalism, which accounts for only the time evolution of the spin degrees of freedom, and the structural properties are obtained for the SC and BCC magnetic systems shown in Fig1.
The Landau-Lifshitz-Gilbert equation has been numerically solved by replacing the spin operators of the Heisenberg Hamiltonian with classical angular momentum vectors. The evolution of each spin can be seen as its precession around an effective field $\mathbf \Omega_i=-\sum_{j\neq i}J_{ij}\mathbf S_j$ induced by the neighboring spins. Furthermore, to induce energy dissipation,  the system is physically embedded in a thermal bath at constant temperature, mathematically represented by a stochastic field and a dissipative term in the equation. 
The differential equation of motion reads
\be\label{eq:spin_dyn}
\frac{d\mathbf S_i}{dt}=\frac{1}{\hbar} \left[\mathbf S_i\times\left(-\mathbf\Omega_i+\mathbf h _i\right)-\gamma\mathbf S_i \times \left(\mathbf S_i \times (-\mathbf \Omega_i)\right)\right]
\ee
where $\mathbf h_{i}$ is a Gaussian distributed white noise with zero mean and vanishing correlator  $<h_{i}(t)h_{j}(t')> = \mu \delta_{i,j}\delta(t,t')$, representing the stochastic field. Its value at each time step $\Delta t$ is $h_{\alpha,i}=\eta\sqrt{\mu/\Delta t}$ with $\alpha \in \lbrace x,y,z \rbrace $ being the cartesian coordinate and $i$ referring to the site label, $\eta \sim  \mathcal N\left(0,1\right)$ is a random variable sampled from the standard normal distribution.
 $\gamma$ is the Gilbert dissipation parameter related to the stochastic field through the fluctuation-dissipation theorem (FDT) \cite{Chandrasekhar_1943, Kubo_1966} $\mu=2\gamma k_B T $ at equilibrium. We point out that $\gamma$ is a dimensionless parameter that can be extracted directly from the spin dynamics simulation or by comparison with Monte Carlo calculations at equilibrium. We found that both are consistent within the error bars (Fig \ref{fig:MC_SD}). The dynamics is implemented via the Suzuki-Trotter decomposition outlined in Refs.\cite{Dudarev_2010,Ma_2011} and time units have been fixed assuming the exchange interaction is of the typical order of magnitude of the bulk $J=0.1$ eV.

\subparagraph{Spin dynamics after a ferromagnetic quench.}
\begin{figure}[!b]
\begin{center}
\includegraphics[width=0.5\textwidth,height=0.4\textwidth]{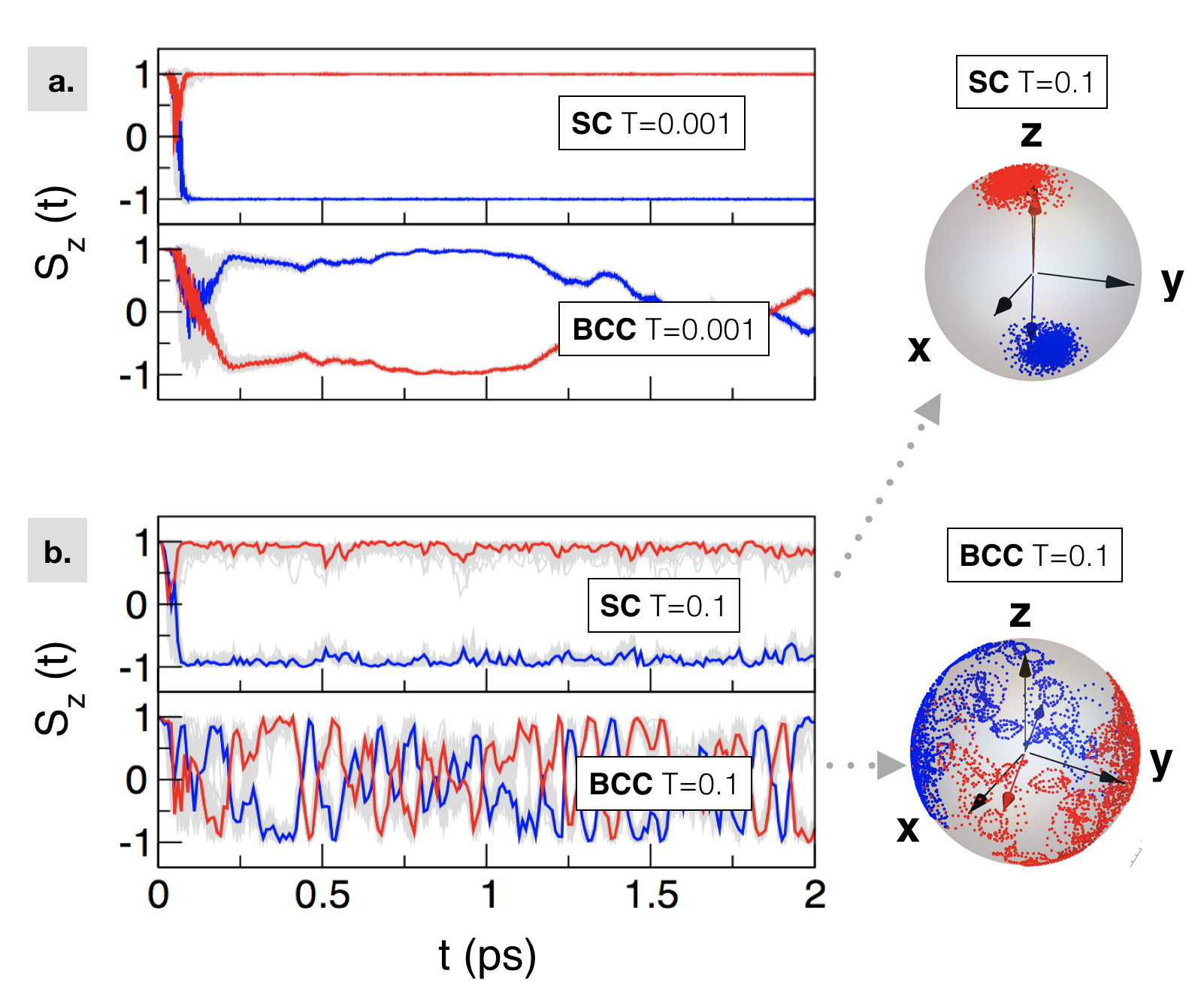}
\caption{a-b)Time evolution of the z-component $S_z(t)$ of all the spins (grey solid lines) in the system initially subjected to a ferromagnetic quench: blue and red solid line refer to two sites with opposite magnetic moments after the transient.  Results are shown for $T=0.001, 0.1 [J_0]$ respectively for SC and BCC.   On the side, the spin configuration at $T=0.1[J_0]$ for two different sites (blue and red) with opposite magnetic moments at the steady state is shown. }\label{fig:dyn_nofield}
\end{center}
\end{figure}
\begin{figure}[!t]
\begin{center}
\includegraphics[width=0.35\textwidth,height=0.33\textwidth]{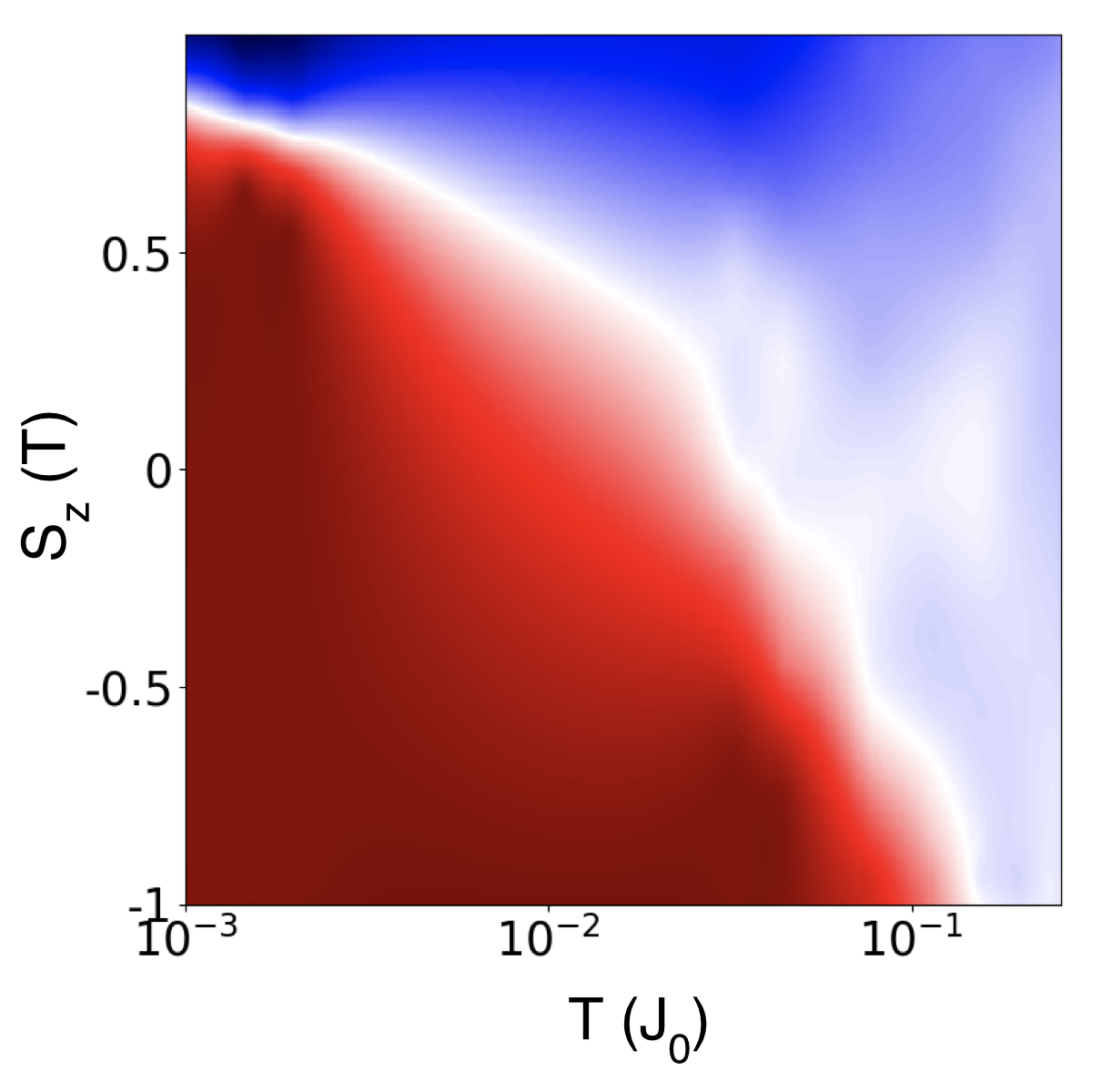}
\caption{Histogram $P(S_z)$ of the z-axis spin component obtained along the time dependent trajectories, at different temperatures $T$ in the SC. The color map goes from zero (red) to one (blue). Dynamics considered up to $t_f=12$ ps. }\label{fig:hysto_memory}
\end{center}
\end{figure}
Since we are interested in the response of the system to an initial out-of-equilibrium condition, we start from the ferromagnetic configuration and let the system evolve.  At low temperature ($T=0.001[J_0]$, Fig \ref{fig:dyn_nofield}a ) the central site of the nanostructures (blue line for SC and red for BCC) anti-aligns with respect to its initial configuration along the z-direction under the effect of the evolution operator. The response of the lattice to this mechanism is prompt: the flipping mechanism diffuses from the core spin to the surface ones ( Fig \ref{fig:radial_dep_evolution} ) and the evolution of all spins is synchronized toward the relaxation to the antiferromagnetic  (AF) state by keeping a global orientation on each of the sublattices. Red and blue lines in Figs. \ref{fig:dyn_nofield}(a) and  \ref{fig:dyn_nofield}(b) represent the evolution of the z-component of two spins belonging  to  ferromagnetic sublattices A and B. The dynamics of the local vector on each site is also shown via the gray shaded area and it shows that the time dynamics is synchronous. 
We note however that, differently from the BCC, at $T=0.001[J_0]$ (Fig3a) in the SC only one sublattice is evolving under the time evolution operator, while the other is not affected by the dynamics, and the orientation of the initially applied ferromagnetic field  $\mathbf B_Q=(0,0,1)$  is retained (red line). This non-trivial dynamics is triggered by the exchange coupling between the sublattices in the AF phase when the system is initially prepared out-of-equilibrium. The spontaneous locked polarization of the magnetization in the SC is due to its intrinsic weak ferromagnetic component. Indeed, even if both SC and BCC are able to stabilize a N\'eel order along all the directions, they differ in the number of uncompensated spins (10\% SC and  0.1\% BCC ).  
Thus while for BCC the flipping mechanism happens at $\approx 2$ ps, for the SC it occurs at $t \approx 0.4$ ns for fixed  $T=0.001 [J_0]$ (see Fig \ref{fig:dyn_longt}a).
We now consider the dynamics at higher temperatures $T=0.1 [J_0]$ (Fig \ref{fig:dyn_nofield}b) where the systems are still magnetic. We observe that while for the BCC ergodicity is recovered, for SC  the obtained time evolution is similar to the $T=0.001[J_0]$ case with thermally activated small oscillations around the initial quenching field $\mathbf {B_Q}$. 
The results discussed so far for short time window, $t_f$=2 ps, are extended up to $t_f$=12 ps. In Fig \ref{fig:hysto_memory} we considered the histogram $P(S_z)$ of the z-axis spin component obtained along  the  time  dependent  trajectories,  at  different  temperatures $T$.  Note  that  for  an ergodic system, the histogram is uniform, as obtained for $T >0.1[J_0]$.  For $T <0.1[J0]$, the spins evolve along a constrained time trajectory, with a non uniform distribution of the spin components, which is a signature of a non-thermal state. We hence observe in the SC long-lived non-thermal states, which are non-trivial topologically protected states driven by the interplay of the initial quench and  long-range interactions. On the other hand in the BCC, whose structure is not bipartite, the intrinsic magnetic anisotropy due to the geometry does not stabilize a long-lived memory effect against thermal fluctuation.

\begin{figure}[!b]
\begin{center}
\includegraphics[width=0.44\textwidth,height=0.5\textwidth]{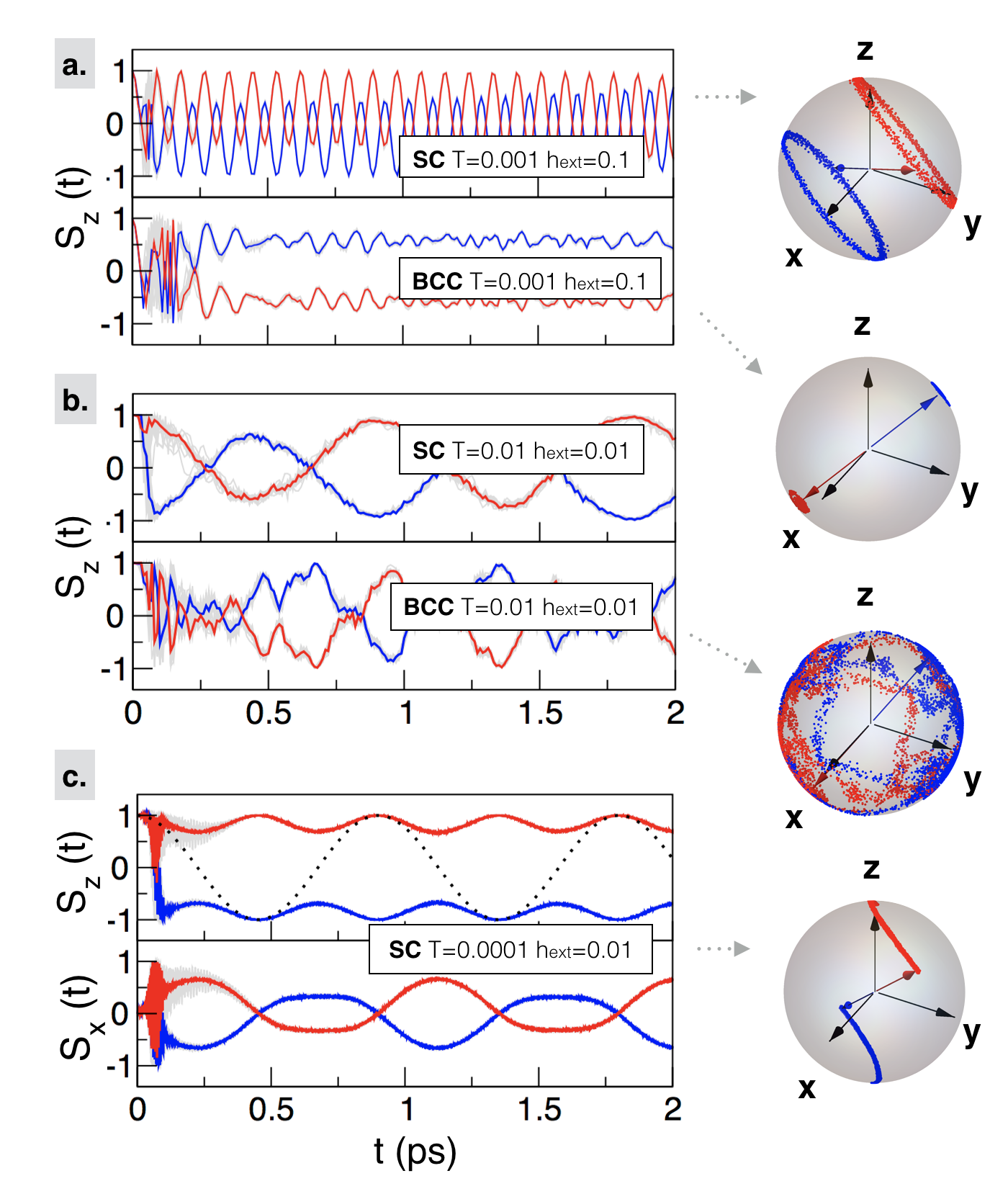}
\caption{a-b) Time evolution of the z-component $S_z(t)$ of all the spins (grey solid lines) in the system initially subjected to a ferromagnetic quench at $t=0$ and a static field for $t>0$. Blue and red solid lines refer to two sites with opposite magnetic moments after the transient.  Results are shown for different temperatures and values of the fields $h_{ext}$ respectively for SC and BCC. c)  Time evolution of the x and z-component of the spins $S(t)$ in the system subjected to a ferromagnetic quench at $t=0$ and an AC field (black dashed lines $\mathbf {h_{AC}}/h_{ext}$) for $t>0$. On the side, the spin configurations for two different sites are shown.}\label{fig:dyn_field}a
\end{center}
\end{figure}

\subparagraph{Adding a static field.}
Interestingly, time translational symmetry observed in the transient in Fig \ref{fig:dyn_nofield}a is explicitly broken when an applied magnetic field $\mathbf h\rightarrow \tilde{\mathbf h}=\mathbf h+ \mathbf {h_\text{ext}}$ with $h_{ext,i=\{x,y,z\}}=h_{ext}$ is switched on (Eq.[2]) after the ferromagnetic quench $\mathbf {B_Q}$. Indeed the spin dynamics after the quench and an applied field shows a periodic behavior. Here we consider a small external field ($h_\text{ext}=0.01, 0.1$) and we observe the dynamics in the temperature regime where the effects of the small field are not washed away by thermal fluctuations. 
Fig \ref{fig:dyn_field}a shows that while for the BCC cluster the staggered magnetic vector stabilizes along the direction of the field, in the SC cluster the dynamics is characterized by  long-lived oscillations and the alignment is recovered at  $t=20$ ps for $T=0.001 [J_0]$, $h_\text{ext}=0.1$ (Fig \ref{fig:dyn_longt}b). Thus in the SC the dynamics after the quench is characterized by two time scales. First, at a femtosecond scale  the antiferromagnetic interaction brings the system toward its energy minimum: the spins after the ferromagnetic quench follow the same relaxation as in absence of applied field at first and form an AF state. Second, at a larger time scale the AF state symmetry is maintained and the spins precess around the $\mathbf {h}_{ext}$ direction: this not trivial dynamics which brings the system towards the alignment along the direction of the external field is characterized by observable timescale (Fig \ref{fig:dyn_longt}b). The arising of this intermediate  phase between the AF stabilization and the alignment with the external field, characterized by a periodic response of the system to a static field,  is exclusively triggered by the uncompensated spins in the AF nanocluster. 
The scenario at $T=0.01$ and $ h_\text{ext}=0.01$ (Fig \ref{fig:dyn_field}b) shows that the periodic phase persists even for higher temperature and smaller field: indeed while in the BCC the thermal fluctuations destroyed the effect of $\mathbf h_\text{ext}$, in the SC the periodic oscillations are still evident. \\
Thus we discover an interesting phase of the transient triggered by the AF interactions in a finite system where the response to a static field shows a periodic behavior with an observable and tunable relaxation time scale. 

\subparagraph{Adding an AC field}

\begin{figure}[b]
\begin{center}
\includegraphics[width=0.38\textwidth,height=0.47\textwidth]{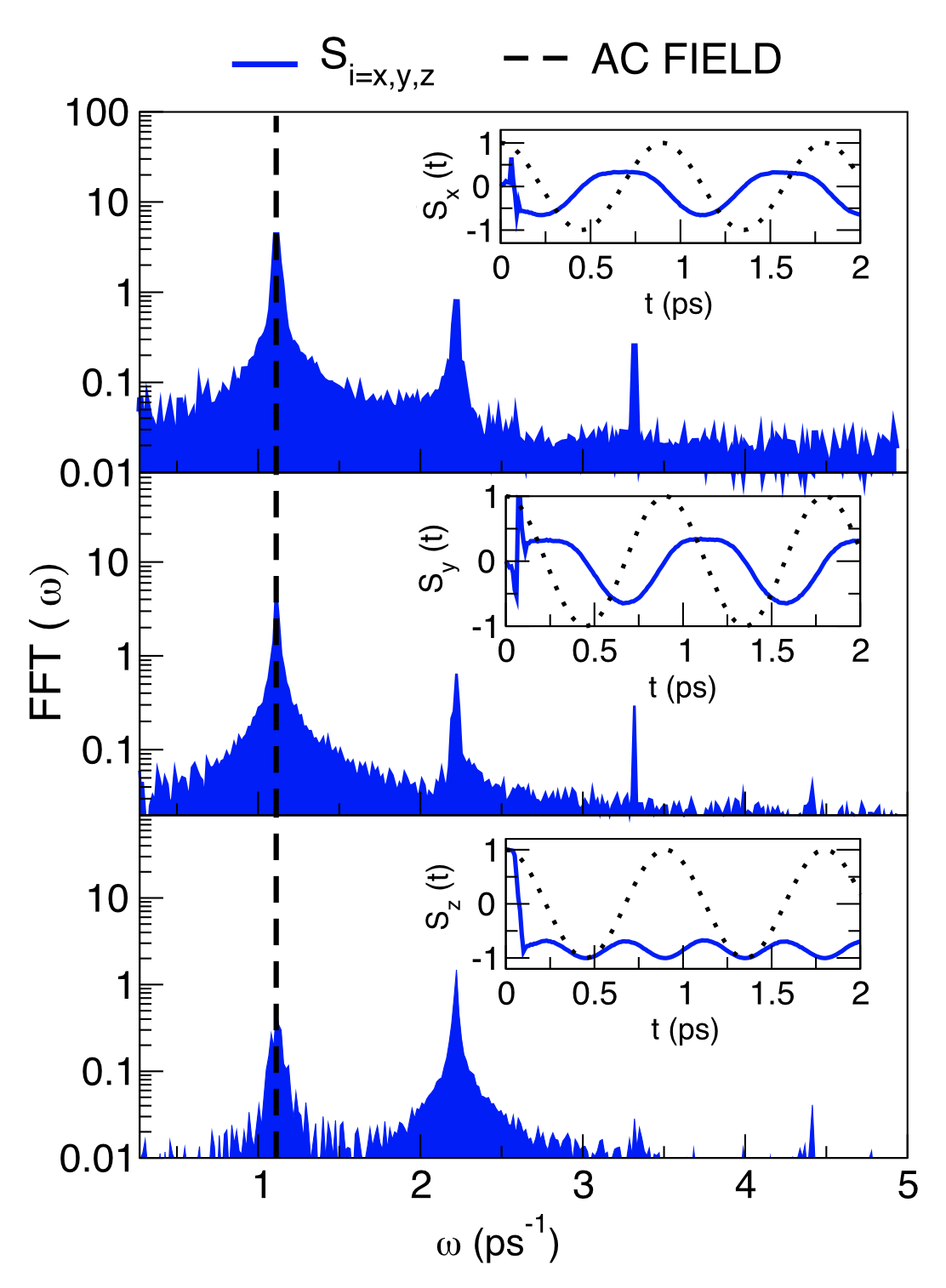}
\caption{ Fourier analysis of the core spin trajectories (blue solid lines) compared with the AC field  $\mathbf{h_{AC}}=\mathbf{h_\text{ext}} \cos(2\pi\omega t)$ (black dashed lines for $\mathbf {h_{AC}}/h_{ext}$). Results are shown for fixed $T=0.0001$ and $h_\text{ext}=0.01$ values. Time window considered after the transient up to $70$ ps. }\label{fig:FT_dyn_ACfield}
\end{center}
\end{figure}

Finally,  we study the response of the SC system to an AC field $\mathbf h\rightarrow \tilde{\mathbf h}=\mathbf h+ \mathbf{h_\text{ext}} \cos{(2\pi\omega t)}$ after the ferromagnetic quench $\mathbf {B_Q}$. 
Fig  \ref{fig:dyn_field}c shows the dynamics of the spins in the time crystal phase. In the latter phase, the time evolution is not periodic with the applied field,  and in particular we find that the component of the spin aligned in the quenched direction (e.g. along z-axis) oscillates with a frequency which is double of the applied field (see Fig \ref{fig:FT_dyn_ACfield}). In our setup, we typically use an applied field in the terahertz range (1.1 THz).
Note that further sub-harmonics are also present with albeit smaller amplitudes.
Thus, the initial ferromagnetic quench and the oscillating applied field prevent the spins from completing a full precession, and instead the covered phase space is limited to a portion of the AF precession circles. The spin trajectories explicitly break the time invariance observed in Fig \ref{fig:dyn_nofield}a, and enter into a periodic motion that lasts beyond the largest time considered in our calculations ($\approx 1$ ns, Fig\ref{fig:dyn_longt}d).

This non-trivial time translational symmetry breaking, which protects the $\mathbb Z_2$ topology of the spin trajectories for macroscopic time scales, bears similarities to the time crystal \cite{Patrick_2013} recently observed experimentally in magnetic systems \cite{Zhang_2017}. In our calculations, the time-crystal phase (TC) stems from non-thermal states triggered by the non-trivial dynamics obtained after the quench,  when, at $T=0.0001 [J_0]$,  only one of the bipartite sub-lattice evolves with time, whereas the other sub-lattice stays aligned with the applied magnetic field $\mathbf {B_Q}$, and remain topologically protected upon application of AC field.  At finite but small temperatures this dynamics breaks the time translational invariance and the spherical symmetry of the spins and in turn provides a nanoscopic time crystal with the magnetic vector aligned along a preferred orientation (induced by a combination of the initial quench and applied magnetic field). Thus in the TC phase we observe non thermal states and violation of the FDT (Fig \ref{fig:FDT_chi}). The  loss  of  the  time  crystal  phase  it  is  in  particular  accelerated increasing the dissipation term $\gamma$ (Eq.\ref{eq:spin_dyn}) (see Fig \ref{fig:dyn_ACfield_details}a) or  the temperature (Fig \ref{fig:dyn_ACfield_details}b), as the thermalisation happens on faster time scales. \\

\paragraph{Conclusions}
In conclusion we provided the results of  the role of the spin-lattice coupling in the equilibration of an antiferromagnetic  nanoscopic system at both zero and finite temperature. We then focused on the spin quenched dynamics in a temperature regime which foregoes the melting of the magnetic moment. 
We demonstrated slow relaxation and non-ergodicity in non-disordered nanoscopic many-body systems induced by the initial magnetic quench. The non-thermal states persist in the presence of experimentally controllable classical thermal noise and the signatures of metastability are uncovered in situations where non-ergodicity is transient only due to dissipation. 
Our work lays out foundations for future experiments in small antiferromagnetic nanoparticles\cite{Mellouhi_2013, Han_2005, Wang_2011}. Furthermore, it provides numerical results for future realisation of antiferromagnetic memory devices in which magnetocrystalline anisotropy \cite{Loth_2012}  and tunability of the exchange coupling \cite{Oberg_2013}  play a crucial role. \\

\paragraph{Acknowledgment} We gratefully acknowledge insightful discussions with S.L. Dudarev and P.-W. Ma. C.L. is supported by the EPSRC Centre for Doctoral Training in Cross-Disciplinary Approaches to Non-Equilibrium Systems (CANES, EP/L015854/1). C.W. gratefully acknowledges the support of NVIDIA Corporation, ARCHER UK National Supercomputing Service. We are grateful to the UK Materials and Molecular Modelling Hub for computational resources, which is partially funded by EPSRC (EP/P020194/1). 

\appendix 
\renewcommand\thefigure{\thesection.\arabic{figure}}    
\renewcommand\thetable{\thesection.\arabic{table}}    
\setcounter{figure}{0}    
\setcounter{table}{0}

\section{Monte Carlo simulation}
As reported in the main text, at zero temperature the competition between the Heisenberg term and the Lennard-Jones potential, respectively tuned by $\alpha$ and $K$ parameters introduced in Eq.\ref{eq:hamiltonian_H_LJ}, stabilizes three different ground state structures (Fig \ref{fig:GS_phase_diagram}): simple-cubic (SC), body-centered-cubic (BCC) and hexagonal-close-packed (HCP). These structures are obtained either if the system optimizes the super-exchange potential or the ionic interactions (LJ). Under the assumption of large magnetic moment, paradigmatic systems for the model above are represented by antiferromagnetic SC structures such as perovskites (e.g. LaTiO$_3$ \cite{Mellouhi_2013, Luan_2015,Weng_2015}), BCC nanoclusters such as Mn \cite {Briere_2002} 
and HCP Ni nanoparticle \cite{Garcia_2011, Filippova_2015, Tranchida_2018}. An estimation of the order of magnitude  of the parameters $(\alpha, K)$ for the examples provided above can be inferred combining two features: i) the structural properties - e.g bond lengths, coordination number, radial distribution function -  of the clusters obtained in our simulation and ii) the model of the exchange interaction and Lennard-Jones potentials used respectively for bulk and nanoclusters available in the literature..\\
A sensible estimation of the $\alpha$ parameter is obtained from the analysis of the exchange interaction for a selected element or compound. In particular we will use the Bethe-Slater equation properly parametrised for a given element/material (e.g. in Refs.\cite{Tranchida_2018, Cardias_2017} ) or the computed nearest- $J_1$ and the next nearest $J_2$ exchange coupling obtained mapping the density functional theory (DFT) energies to a classical Heisenberg model (Ref \cite{Weng_2015}). $\alpha$ is obtained in units of \AA $^{-1}$ .
The LJ interaction is extracted from previous works and molecular dynamics  calculations carried out in Ref.\cite{Filippova_2015, Duarte_2014,Luan_2015}. $K$ is obtained in units of eV.\\ 
We report in Table \ref{tab:table_LJ_H} the inferred values of $(K,\alpha)$ for the clusters under analysis, properly supported by several references. 
\textcolor{black}{We emphasize however that such a correspondence between values obtained by \textit{ab initio} DFT and the Heisenberg theory remains at the qualitative level, as quantum fluctuations are not included for classical spin systems.}
\begin{table}[htp]
\begin{center}
\begin{tabular}{c|c|c}
&$K$ [eV] &$\alpha$ [\AA $^{-1}$]\\
\hline 
\hline
\textbf{Ni} 	\quad\quad	& 	\quad\quad0.66 \cite{Filippova_2015}\quad \quad	&	\quad\quad  2.18  \cite{Tranchida_2018}\\
\hline
\textbf{Mn} \quad\quad	& 	\quad\quad0.55 \cite{Duarte_2014} 	\quad\quad	& 	\quad\quad 7.84 \cite{Cardias_2017}\\
\hline
La\textbf{Ti}O$_3$ \quad\quad& \quad\quad0.02 \cite{Luan_2015}      \quad\quad      & \quad\quad 2.79 \cite{Weng_2015}\\
\end{tabular}\label{tab:table_LJ_H}
\end{center}
\caption{$(K,\alpha)$ parameters inferred for the three paradigmatic systems provided as examples for the magneto-structural phases in Fig \ref{fig:GS_phase_diagram}.  }\label{tab:table_LJ_H}
\end{table}%


\section{Spin Dynamics simulation}

\subsection{Benchmark of stochastic time dependent calculations against Monte Carlo simulations}
\begin{figure}[!t]
\begin{center}
\includegraphics[width=0.4\textwidth,height=0.33\textwidth]{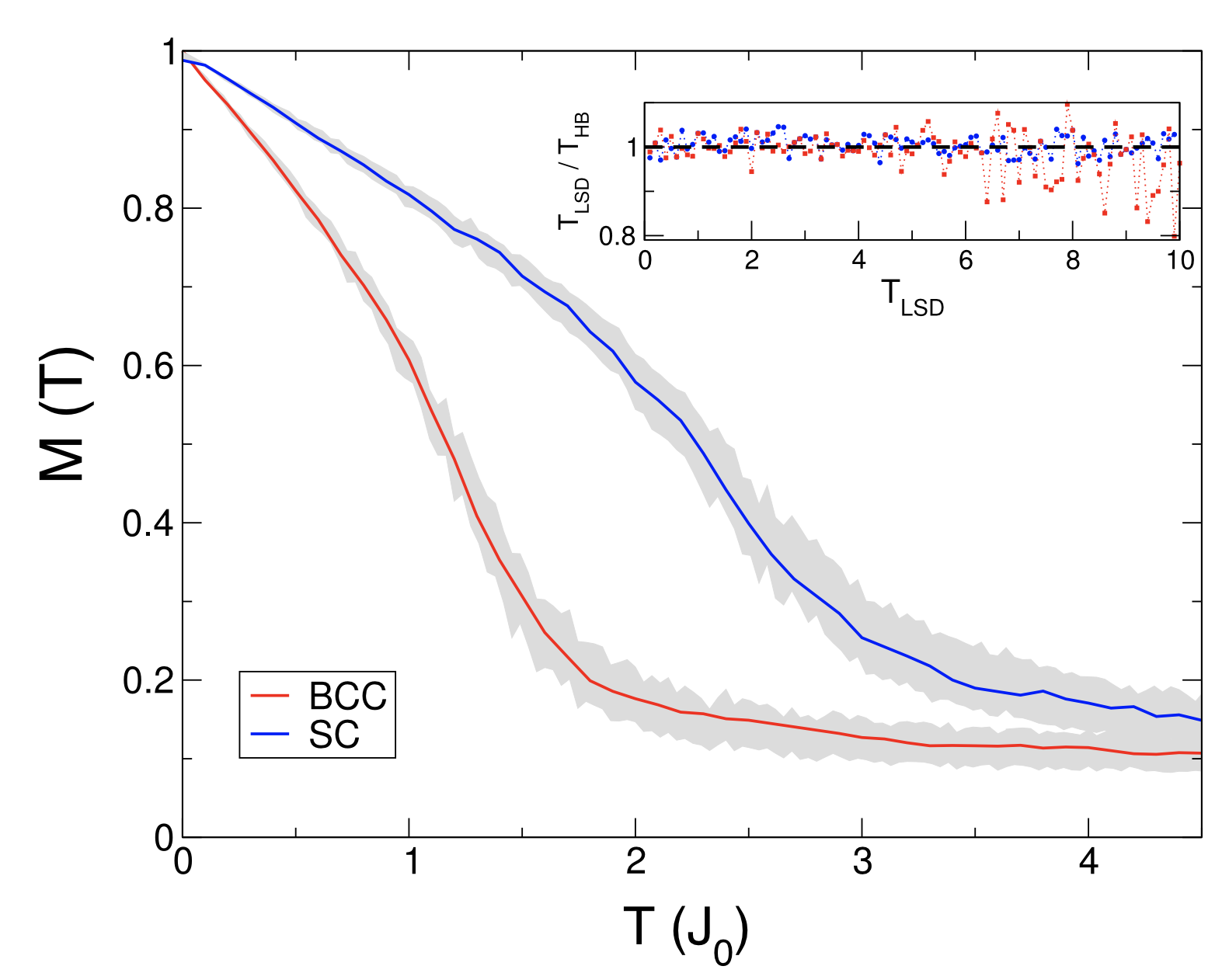}\textit{ab initio}
\caption{Magnetic order parameter as a  function of temperature obtained for two different systems (Simple Cubic (SC) in blue and Body Centered Cubic (BCC) in red). The grey shaded regions represent the Langevin statistics. The inset shows the match between the temperature  obtained with the \textit{Langevin Spin Dynamics } ($T_\text{LSD}$) and the Monte Carlo \textit{heat bath} ($T_\text{HB}$) results.} \label{fig:MC_SD}
\end{center}
\end{figure}

Insights on the dynamical magnetic properties of the nano-structures stabilized so far at equilibrium, are obtained by running spin dynamics simulations.
The Landau-Lifshitz-Gilbert equation in  Eq.\ref{eq:spin_dyn} was numerically solved by replacing the spin operators of the Heisenberg Hamiltonian with classical angular momentum vectors. 
 $\gamma$ is the Gilbert dissipation parameter related to the stochastic field through the fluctuation dissipation theorem (FDT) \cite{Chandrasekhar_1943, Kubo_1966}  $\mu=2\gamma k_B T $ at equilibrium. We point out that $\gamma$ is a dimensionless parameter that can be extracted directly from the spin dynamics simulation or by comparison with Monte Carlo calculations at equilibrium. In Fig \ref{fig:MC_SD}, we show the magnetic order parameter obtained with \textit{heatbath} simulations and the spin dynamics.  Thus the temperature obtained under the hypothesis that FDT is satisfied during the dynamics of the spins (Ref. \cite{Dudarev_2010}) can also be obtained by  the equilibrium Monte Carlo simulation.
 
\begin{figure}[!t]
\begin{center}
\includegraphics[width=0.38\textwidth,height=0.38\textwidth]{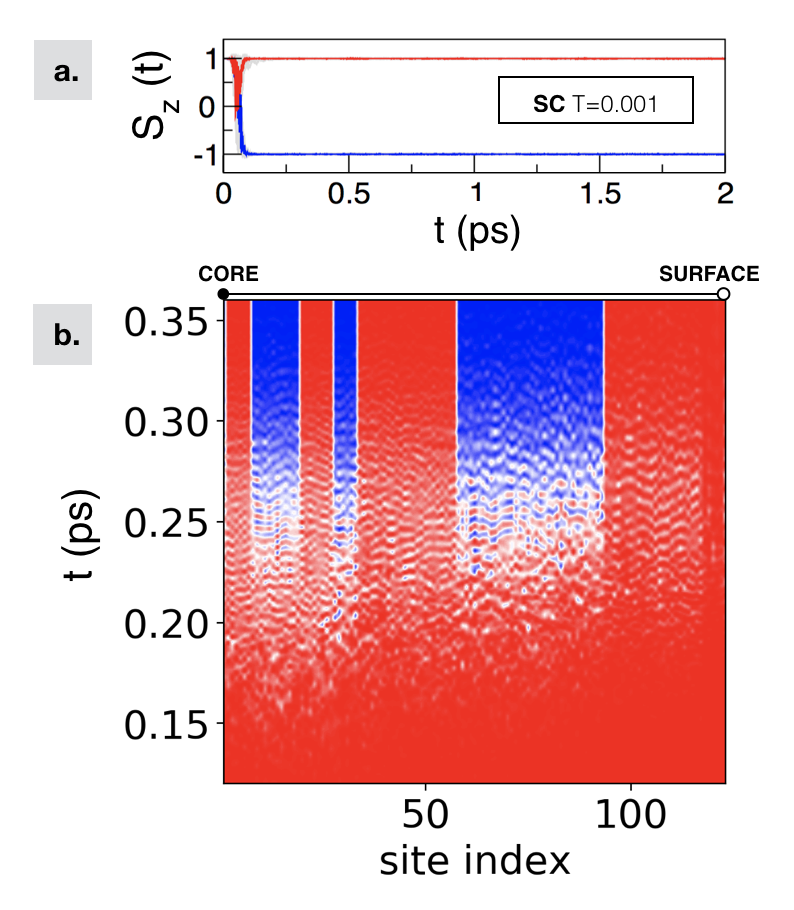}
\caption{ a) Time evolution of the z-component $S_z(t)$ of all the spins (grey solid lines) in the system initially subjected to a ferromagnetic quench: blue and red solid lines refer to two sites with opposite magnetic moments after the transient.  Results are shown for $T=0.001 [J_0]$  for SC nanocluster.  b) Details of the flipping mechanism during the transient at $ T=0.001[J_0]$, which starts from the core site toward the surface. Sites are labeled according to their distance from the center.}\label{fig:radial_dep_evolution}
\end{center}
\end{figure}

\begin{figure}[!h]
\begin{center}
\includegraphics[width=0.40\textwidth,height=0.34\textwidth]{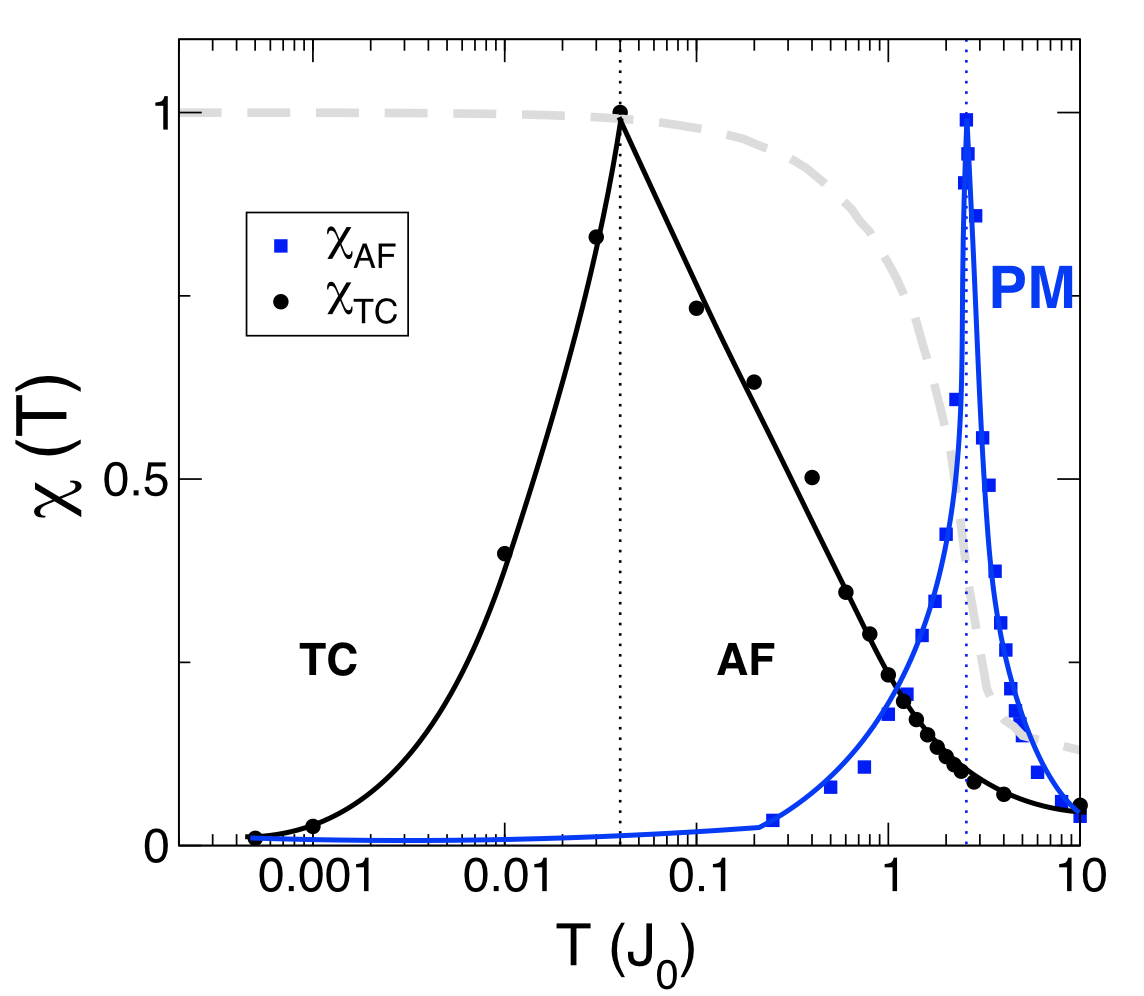}
\caption{ Temperature dependent behavior of the fluctuations of the order parameter both along the time evolution after the quench (black circles) and at the equilibrium one (blue squares). The plot highlights the arising of three different regions: time crystal (TC) , magnetically ordered phase (antiferromagnetic AF) and the paramagnetic phase (PM). The structure under analysis is the SC. The gray dashed line is the staggered magnetic order parameter squared. }\label{fig:FDT_chi}
\end{center}
\end{figure}
\subsection{Insights on the spin dynamics results}

\paragraph{Spin dynamics after a ferromagnetic quench: BCC versus SC nanoclusters}

\begin{figure}[]
\begin{center}
\includegraphics[width=0.4\textwidth,height=0.5\textwidth]{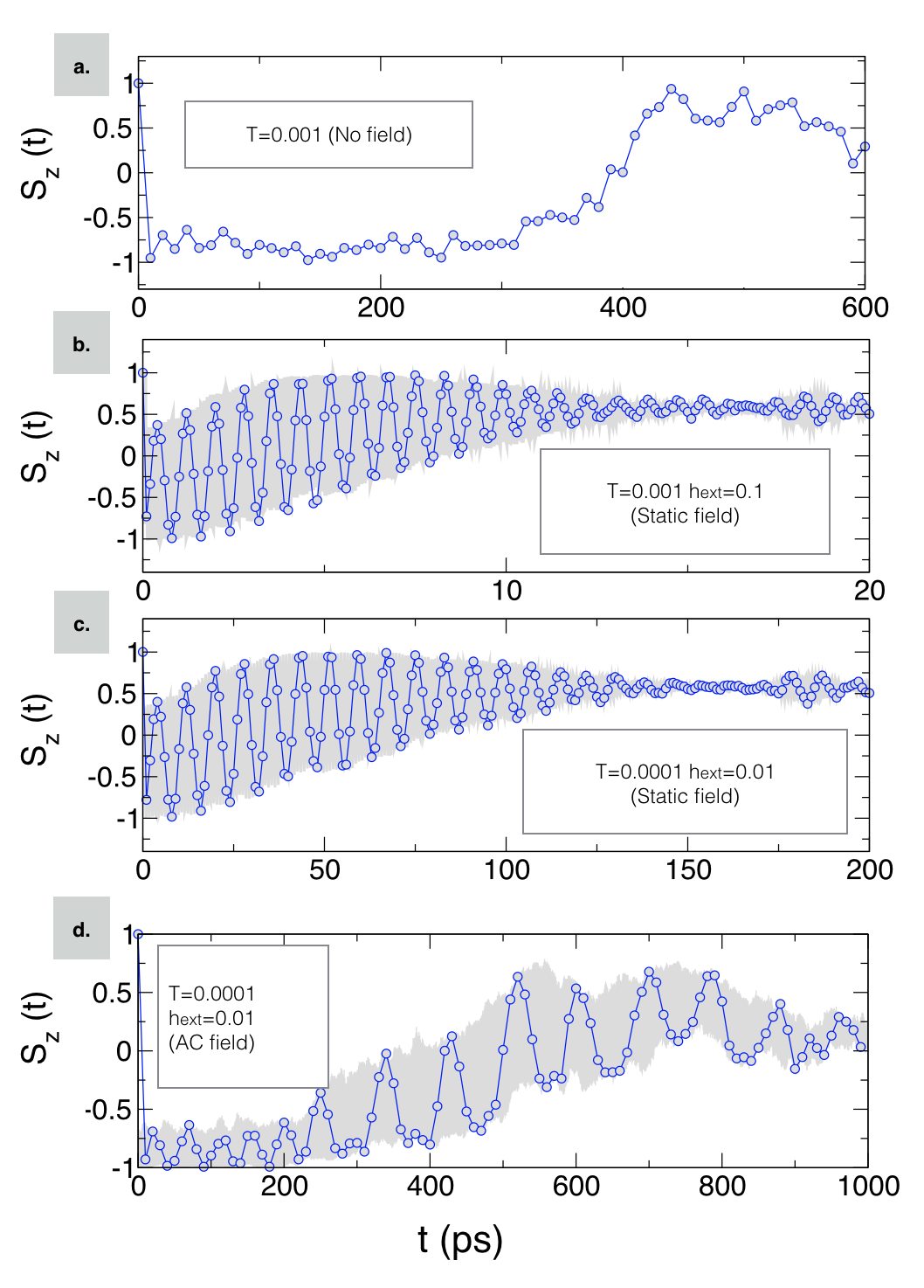}
\caption{a) Time evolution of the z-component $S_z(t)$ of the core spin (blue dotted line) in the antiferromagnetic SC subjected to three different protocols. a) System initially $(t=0)$ subjected to a ferromagnetic quench and subsequently let to evolve in absence of any external drive (long time dynamics of Fig \ref{fig:dyn_nofield}a in the main text); b-c) Ferromagnetic quench (t=0) and a static field $\mathbf{h_{S}}=\mathbf{h_{ext}}$ at $t>0$ (long time dynamics of Fig \ref{fig:dyn_field}a in the main text); d) Ferromagnetic quench $(t=0)$  and  an AC field $\mathbf{h_{AC}}=\mathbf{h_{ext}} \cos(2\pi\omega t)$ (long time dynamics of Fig \ref{fig:dyn_field}c in the main text). Results are shown for different temperatures and $h_{ext}$ values.}\label{fig:dyn_longt}
\end{center}
\end{figure}

We now turn the discussion to the the response of the system to an initial out-of-equilibrium condition. Thus we initialize the system in the ferromagnetic configuration along the z-axis and we let the system evolve with the exponential time evolution operator for the spins given by Eq.(\ref{eq:hamiltonian_H_LJ}). 
As we discussed in the main text, at low temperature ($T=0.001[J_0]$, Fig \ref{fig:radial_dep_evolution}a) the central site of the nanostructures (blue solid line) anti-aligns with respect to its initial configuration along the z-direction under the effect of the evolution operator. The time dependence of the local vector on each site is also shown in Fig \ref{fig:radial_dep_evolution}b where the sites are labeled according to their distance from the center and the color map refers to the value of the z-component for each spin $S_z(t)$ from $+1$ (red) to $-1$ (blue). We note that the flipping mechanism during the transient at $T = 0.001[J_0]$ starts from the central site toward the surface and the evolution of all spins is synchronized toward the relaxation to the AF state by keeping a global orientation on each of the sublattices.
As we observed already in the main text of this work, the SC nano-structure shows a locked direction of the spin along z-axis up to $t \approx 0.4$ ns for fixed  $T=0.001 [J_0]$ (See Fig \ref{fig:dyn_longt}a).

To complete the understanding of this phase with spontaneous locked polarization of the magnetization in the SC cluster, we considered  in Fig. \ref{fig:hysto_memory} the histogram $P(S_z)$ of the z-axis spin component obtained along the time dependent trajectories, for a calculation extended up to $t_f$=12 ps. 
 Note that for an ergodic system, the histogram is uniform, as obtained for $T>0.1 J_0$. For $T<0.1 J_0$, the spins are evolving along a constrained time trajectory, with a non uniform distribution of the spin components, which is a signature of a non thermal state. \\
We now want to look at the average quantities which characterize  the dynamics of the structures considered and we use the fluctuation dissipation theorem  (FDT) to search for the evidence of the non-thermal states observed above. In the following analysis we focus on the SC for which, due to its bipartite structure, the memory effect is enhanced and resistant against thermal fluctuation even in absence of magnetic anisotropy. 
In Fig \ref{fig:FDT_chi} we calculate both the correlation of the order parameter along the time evolution after the quench and its fluctuations at equilibrium as $ \chi_{\mathbf q=\text{AF}}(T)$ (where the wave vector dependent magnetization $M_\mathbf q=(1/N)\sum_i\mathbf S_i e^{i\mathbf q\cdot\mathbf r_i}$ has $\mathbf q$ corresponding to ordering wavevector defined for the bulk antiferromagnetic states $\mathbf q=(\pi,\pi,\pi) / a$ for SC).
At equilibrium, the two quantities are related by the FDT. However out-of-equilibrium, in a case of a non-ergodic dynamics, a violation of the FDT theorem can be observed for non thermal state.
We observe indeed in our calculations the presence of long-lived non-thermal states obtained after the quench, violating the FDT, at temperature $T<0.01 [J_0] $ for SC (and $T<0.0001 [J_0] $ for BCC). 
We hence found three different phases: i) non ergodic but magnetic (TC), ii) ergodic and magnetic (AF) and iii) paramagnetic (PM).
In particular the first phase is named a time crystal phase because of time translational invariance

\begin{figure}[b]
\begin{center}
\includegraphics[width=0.35\textwidth,height=0.42\textwidth]{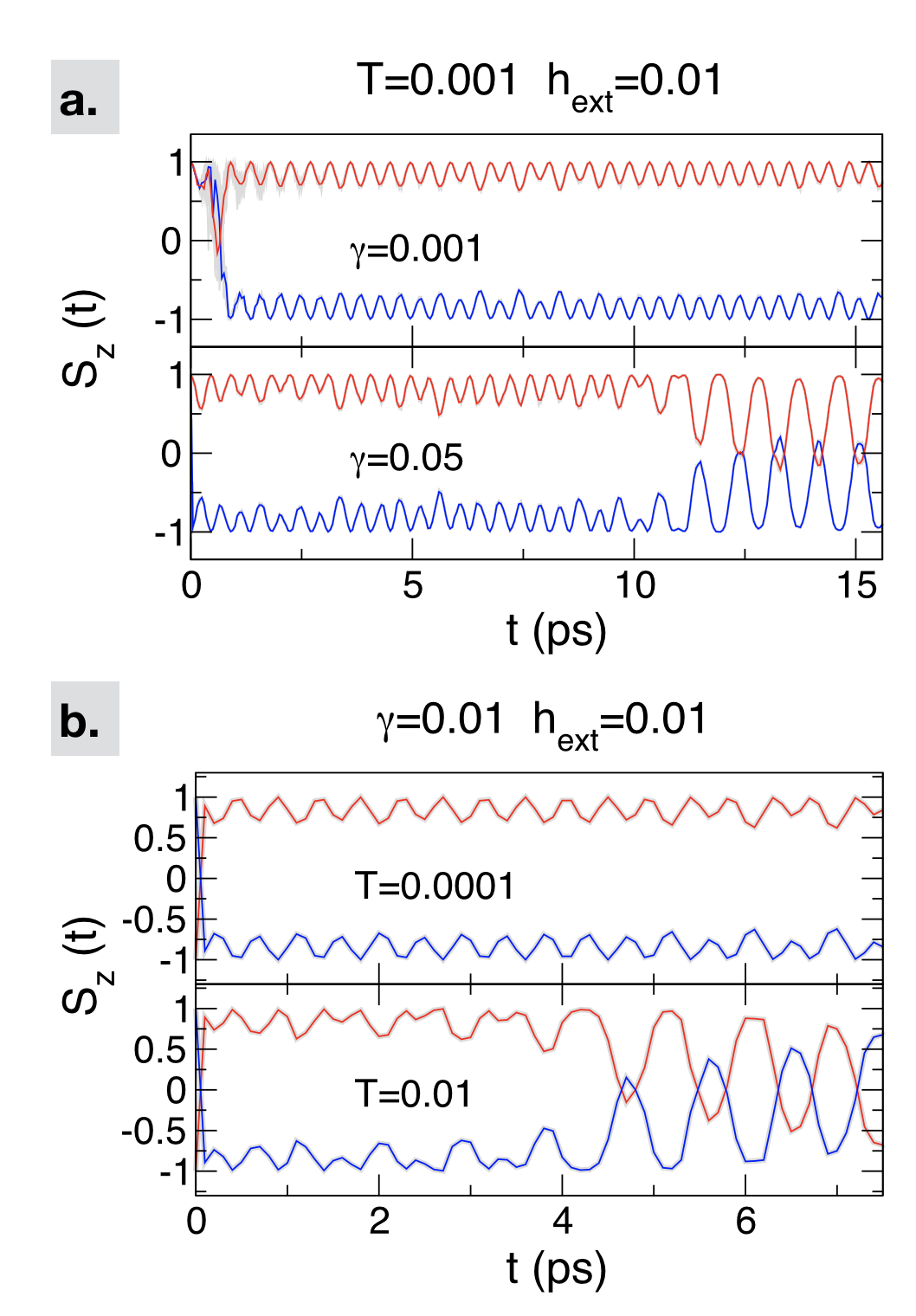}
\caption{ Time evolution $S_z(t)$ in the SC cluster subjected to a ferromagnetic quench at $t=0$ and an AC field $\mathbf{h_{AC}}=\mathbf{h_\text{ext}} \cos(2\pi\omega t)$ for $t>0$. Blue and red solid line refer to two sites having opposite magnetic moments after the transient.  Results are shown for different values of the dissipation parameter and temperature. }\label{fig:dyn_ACfield_details}
\end{center}
\end{figure}

\paragraph{Long time dynamics under the effect of different driving fields}
The results discussed so far for a short time window ($t_f$=2 ps) are extended up to $t_f=$1 ns in Fig \ref{fig:dyn_longt}, where we considered the time evolution of the z-component $S_z(t)$ of the core spin (blue dotted line) in the antiferromagnetic SC subjected to a ferromagnetic quench at $t=0$ and  three different protocols at $t>0$:  i) absence of any external drive in Fig \ref{fig:dyn_longt}a,
ii) introduction of a  isotropic static field $\mathbf{h_S}=\mathbf{h_{ext}}$ at $t>0$ in Fig \ref{fig:dyn_longt}(b-c) and iii) an AC field  $\mathbf{h_{AC}}=\mathbf{h_{ext}} \cos(2\pi\omega t)$ in Fig \ref{fig:dyn_longt}d. In our results we considered small external fields ($h_{ext}=0.1,0.01$) and we observe the dynamics in the temperature regime where the effects of the small field are not washed away by thermal fluctuations.
As we discussed before,  the locked dynamics shown in Fig \ref{fig:dyn_nofield}a after the ferromagnetic quench lasts up to 400 ps as shown in Fig \ref{fig:dyn_longt}a. 
We now want to look at the effect of introducing driving field to this long-lived non-thermal state.\\

In Fig \ref{fig:dyn_longt}b the introduction of a static field $\mathbf {h_S}=\mathbf{h_\text{ext}}$, with $h_{ext,i=\{x,y,z\}}=h_{ext}$,  pointing along a different direction from the quenching field $\mathbf {B_Q}=(0,0,1)$, breaks the translational symmetry observed before. Indeed the spin dynamics after the quench, and upon an applied field, shows a periodic behavior. In particular the dynamics is characterized by two  time scales.  Firstly, at a femtosecond scale  the antiferromagnetic interaction brings the system toward its energy minimum: the spins after the ferromagnetic quench follow the same relaxation as in absence of applied field at first and form an AF state. Secondly at a larger time scale the AF state symmetry is maintained and the spins precess around the $\mathbf {h}_{ext}$ direction. \textcolor{black}{In particular, as shown in Fig \ref{fig:dyn_field}a of the main text, the spins trajectories of the A and B sublattices are locked in one limiting circle (red, aligned with the field) or the other (blue, anti aligned with the field ) giving rise to a $\mathbb{Z}_2$ symmetry of the trajectories. Since the flipping between them is prevented by the interplay of out of equilibrium conditions and the uncompensated spins, we conclude that the $\mathbb{Z}_2$ topology of the trajectories is protected. } As shown in Fig \ref{fig:dyn_longt}b, this non trivial dynamics which brings the system towards the alignment along the direction of the external field is characterized by observable timescale. Thus while for the BCC the staggered magnetic vector stabilizes along the direction of the field at $t=0.5$ ps (main text Fig \ref{fig:dyn_field}a ), in the SC cluster the dynamics is characterized by  long-lived oscillations and the alignment is recovered at  $t=20$ ps for $T=0.001 [J_0]$, $h_\text{ext}=0.1$.  The arising of this intermediate  phase between the AF stabilization and the alignment with the external field, characterized by a periodic response of the system to a static field, is exclusively triggered by the uncompensated spins in the AF nanocluster. The duration of the precession phase is increased decreasing the dissipation term $\gamma$ Fig \ref{fig:dyn_ACfield_details} (a) or the termperature Fig \ref{fig:dyn_ACfield_details} (b).\\

\bibliographystyle{ieeetr}

\end{document}